%% file: paper.tex
\documentstyle[12pt,a41,cite]{article}
\newcommand{\myappendix}{\setcounter{equation}{0}\appendix}
\newcommand{\Li}{{\rm Li}}

\newcommand{\si}{{\rm sign}}

\newcommand{\bq}{\begin{equation}}
\newcommand{\eq}{\end{equation}}
\newcommand\beq{\begin{equation}}
\newcommand\eeq{\end{equation}}
\newcommand\bea{\begin{eqnarray}}
\newcommand\eea{\end{eqnarray}}

\newcommand\Mvec{\,\mbox{\bf M}}

\include{epsfig}

\begin{document}
\noindent
\sloppy

\thispagestyle{empty}
\begin{flushleft}
DESY 05--002 \hfill
{\tt hep-ph/0501178}\\
SFB/CPP--05--02\\
January 2005
\end{flushleft}
%
\vspace*{\fill}
\begin{center}
{\Large\bf Mellin Moments of the Next-to-next-to Leading Order}

\vspace{2mm}
{\Large\bf  Coefficient Functions for the Drell-Yan Process}

\vspace{2mm}
{\Large\bf and Hadronic Higgs-Boson Production}

\vspace{2cm}
\large
Johannes Bl\"umlein$^a$  and
Vajravelu  Ravindran$^{a,b}$
\\
\vspace{2em}
\normalsize
{\it $^a$~Deutsches Elektronen--Synchrotron, DESY,\\
Platanenallee 6, D--15738 Zeuthen, Germany}
\\

\vspace{2mm}
{\it $^b$~Harish--Chandra Research Institute, Chhatnag Road,\\
 Jhunsi, Allahabad, India.}
\\
\vspace{2em}
\end{center}
\vspace*{\fill}
%
\begin{abstract}
\noindent
We calculate the Mellin moments of the next-to-next-to leading order coefficient 
functions for the Drell--Yan and Higgs production cross sections. The results 
can be expressed in terms of multiple finite harmonic sums of maximal weight
{\sf w = 4}. Using algebraic and structural relations between harmonic sums 
one finds that besides the single harmonic sums only five basic sums and their
derivatives w.r.t. the summation index contribute. This representation reduces
the large complexity being present in $x$--space calculations and is well suited 
for fast numerical implementations.  
\end{abstract}
\vspace*{\fill}
\newpage
\section{Introduction}

\vspace{1mm}\noindent
The {\sf Principle of Simplicity}~\footnote{\sc Pluralitas non est ponenda 
sine neccesitate.} is one of the guiding principles in physics~\cite{OCCAM}. 
Whenever possible one seeks for as simple as possible expressions, not 
only to obtain the result in a more compact form, but also to reveal the 
basic structures behind. This applies also to complex computations in 
particle physics. Without achieving suitable simplifications it is often 
impossible to undertake even more involved calculations, such as one order 
higher in the coupling constant in perturbation theory.  

In the present paper we seek for a simplification of the 2--loop 
coefficient 
functions for the Drell--Yan (DY)  process both for unpolarized and 
polarized 
nucleons, and associated to it, to those for hadronic scalar and 
pseudo--scalar 
Higgs boson production in the heavy--mass limit.
\footnote{For the calculation of the Higgs and pseudo-scalar Higgs 
production cross section including mass effects to NLO see 
\cite{HMAS} and the recent reviews \cite{ABDEL}.}
The pioneering 2--loop
calculations\footnote{For NLO results in the case of Higgs--production
see \cite{DAWS}.}
in this field \cite{DY,HGS} were carried out during the last 15 
years and paved the way to understand single--scale quantities in 
Quantum Chromodynamics (QCD), beyond the level of the pole terms. 
They form one of the milestone in the history of QCD. 
The calculations were performed in $x$--space using the QCD--improved 
parton model. Here $x$ denotes the fraction of the momentum of a radiated 
particle to that of the source particle. For the representation of the 
Wilson coefficients a set of up to 77 functions is needed, not counting 
those, which lead structurally to the same Mellin transform~\cite{JK}.    

With the present collider Tevatron \cite{TEV} and the upcoming
large hadron collider, LHC \cite{LHC}, at CERN, the need for precise
predictions from  theory has become more and more important.
These experiments are aimed at not only to discover new particles
but also to measure various parameters of the Standard Model at higher precision 
to confirm the predictions of the theory.  The Higgs boson of the 
Standard Model has  yet to be discovered at these machines. 
Needless to mention that one also hopes to discover various new particles 
predicted in scenarios beyond the Standard Model, as for example supersymmetric particles
or leptoquarks.  The measurement of the  masses and couplings of newly discovered particles 
likewise the known particles require the precise knowledge of the QCD corrections.

On the theoretical side the relevant observables have to be calculated at high accuracy.
The QCD corrections have a considerable
impact both in the discovery channels as well as for the precision 
measurements. Theoretical uncertainties emerge  
from ultraviolet (UV) renormalization and mass factorization scale 
dependence.
These uncertainties can only be diminished extending the calculations to higher orders.

In the differential and total production cross sections for the Drell--Yan process 
as well as for 
Higgs production at hadron colliders the UV--renormalization 
scale dependence is due to the strong coupling constant.  
On the other hand, processes like Higgs production are associated to 
heavy quark final states such as $t \bar t,b\bar b$, or inclusive 
heavy quark productions. They receive a UV--renormalization scale
dependence also from  heavy quark mass effects.

A second source of uncertainty is due to  mass factorization.
Both the partonic cross sections
and the non--perturbative parton densities depend on the factorization scale. 
Since the partonic cross sections are computable order by order
in perturbation theory, their dependence on the factorization scale
is known completely.  Similarly, one can fully determine
the factorization scale dependence of the parton densities
through solving the associated renormalization group equations (RGE) which 
govern their evolution. To diminish both effects significantly usually   
next-to-next-to leading order (NNLO) QCD--corrections have to be performed.

In recent years, there have been several significant developments
in achieving very precise theoretical results to match with the accuracy 
of
the experiments. With the advent of the NNLO non--singlet and singlet anomalous dimensions
\cite{anNNLO} NNLO predictions for the Drell--Yan \cite{DY} and Higgs--production cross 
sections \cite{HGS} are possible. To get some feeling for the theoretical uncertainty coming 
from the scale--dependence
we consider the following two processes: the total Higgs production 
cross section and the cross section for Higgs production associated with a 
top anti-top quark pair.
The total Higgs production cross section is one of the most 
important processes at LHC from the discovery point of view.  This 
production
cross section receives dominant contributions from  gluon--gluon fusion  
through the top quark loop.  Since the leading order correction is of
$O(\alpha_s^2)$ and also involves two gluon densities in the initial state,
the NLO correction is inevitable to reduce the scale
uncertainty.  Unfortunately the NLO correction
is about $80-100\%$ larger than the LO correction, casting doubt on the reliability of
perturbation theory.  Interestingly, the   NNLO corrections
to the process not only give smaller contributions but also reduce
the scale uncertainty significantly, which improves the stability
of the perturbative result.  Similarly, Higgs production associated with
a top anti--top pair suffers from huge scale uncertainties of the order of 
$100-200\%$ at the LO level. The NLO correction  stabilizes the result 
significantly.  The Drell--Yan cross section, which 
is known upto NNLO  \cite{DY},
not only tests the reliability of perturbative QCD but 
also reduces the uncertainties coming from theory in order to make
background studies more reliable for new particle searches
and physics beyond the Standard Model.  

The NLO and NNLO results are technically complicated and they result
in large expressions involving a large number of functions. 
For example the perturbativly computable coefficient functions 
at NNLO for the Drell--Yan or Higgs total cross section depend on nearly 
80 individual functions.  
These belong to the class of Nielsen integrals \cite{nielsen} with a variety of
partly complicated arguments and products with logarithms or the weights $1/(1\pm x)$.
For the physical cross sections 
these  are convoluted with the respective evolved NNLO parton densities.
Since the different sets of parton density are usually available in form of numerical codes, 
the Mellin convolution is done performing the respective integrals numerically.
Because of the complexity of the expressions involved, the numerical computations 
are usually slow.  

We will discuss here an alternative method which is
faster to compute the cross sections.  It involves the use of the
Mellin transformation technique both for the Wilson coefficients and the   
parton densities. The evolution equations of the parton densities can be 
solved analytically 
in Mellin--space. If the respective Wilson coefficients are also known in Mellin--space
one may calculate the cross sections by a single numerical integral around the singularities
of the product of both in the complex plane. This requires the Mellin--representation of the
Wilson coefficients in the variable $N$ and their analytic continuation to complex values of 
$N$ \cite{JB3}.\footnote{The analytic continuations of the basic Mellin 
transforms which contribute to the 3--loop anomalous dimensions have been 
given in \cite{JBSM} recently.}
 For integer values of $N$   
the Mellin transformation of the Wilson coefficients leads to finite 
harmonic sums \cite{JK,JVE}. We use various algebraic relations between these sums 
\cite{JK,euler,GIRG,JB1} to simplify the expressions. Furthermore, structural relations 
between the Mellin transforms \cite{JB04} are used to reduce the number of 
basic functions further.
The use of finite harmonic leads to a synchronization of the expressions. Several complicated 
harmonic sums present for individual contributions disappear in the final result.
A brief summary of the results presented below was given in 
\cite{JBVR1}.
Similar investigations as performed in the present paper are carried out for the Wilson 
coefficients for deeply inelastic scattering off unpolarized and polarized targets 
\cite{JBSM1} and the unpolarized and polarized time--like fragmentation functions 
\cite{JBVR2} to NNLO. 

The paper is organized as follows. In section~2  we summarize aspects of the 
calculation of 2--loop Wilson coefficients in $x$--space. The Mellin moments are 
discussed in section~3. In section~4 a brief survey is given on multiple 
harmonic sums and 
their relation to Mellin transforms. The algebraic relations between the finite harmonic 
sums used in  the present paper is discussed in section~5. In section~6 
the basic functions 
representing the 2--loop Wilson coefficients for the processes considered in the present 
paper are identified. In the appendices A.1,2 the explicit results are summarized for the 
Mellin transforms of the NNLO Wilson coefficients for the unpolarized and polarized 
Drell--Yan process and appendix A.3,4 contains the corresponding expressions for the 
hadronic scalar and pseudo--scalar Higgs boson production processes.  
\section{Coefficient Functions}

\vspace{1mm}\noindent
Due to mass factorization, hadronic cross sections such as those for the
Drell--Yan process \cite{DY} and hadronic Higgs boson  production 
\cite{HGS} can be expressed 
in terms of Mellin 
convolutions
of the perturbativly computable coefficient functions
$\Delta_{ab}(x,Q^2,\mu^2)$ and 
non--perturbative parton distributions, $f_a(x,\mu^2)$, of incoming hadrons,  
\begin{eqnarray}
\sigma(x,Q^2)&\!\!\!=\!\!\!&\int_0^1 \!\!{dx_1 }\! \int_0^1 
\!\!{dx_2}
\int_0^1 dz
f_a(x_1,\mu^2) f_b(x_2,\mu^2)
\Delta_{ab}\Bigg(z,
{Q^2 \over \mu^2}\Bigg) \delta(x-z x_1 x_2)~.
\label{eq1}
\end{eqnarray}
Here the sum over $a,b={q,\bar q,g}$ is implied.  $Q^2$ is
the mass squared of the Drell--Yan pair or the Higgs boson.  The 
parameter
$x$ is a scaling variable defined by
\begin{eqnarray}
x={Q^2 \over S},\quad \quad \quad  S=(P_1+P_2)^2,
\end{eqnarray}
where $P_1,P_2$ are the momenta of incoming hadrons.  We have set
both the UV-- and factorization scales to be equal to $\mu^2$.
Though the parton densities are not calculable in the perturbative QCD,
their evolution with respect to the factorization scale $\mu^2$ is computable  
using the renormalization group equations (RGE).
Hence higher order corrections to hadronic reactions enter
through two sources viz, the coefficient functions and 
the RG equations of the parton distribution functions.

The coefficient functions are computed from the partonic cross sections
in powers of the strong coupling constant $\alpha_s(\mu^2)$ in perturbative QCD.  
They are expressible in terms of the
scaling variable $z=Q^2/s$, where
$s$ is the center of mass of incoming partonic system.
$Q^2$ is a large invariant mass characteristic for the process.
The lowest order contributions are of $O(\alpha_s^0)$ and the Wilson coefficients 
are obtained from the Born--diagrams.
In the case of the Drell--Yan process the lowest order contributions are due to quark 
anti-quark annihilation, likewise gluon--gluon fusion in the case of Higgs boson 
production through the top--quark loop.  Hence the results are just proportional 
to $\delta(1-z)$.  To the next-to-leading order (NLO) 
one encounters virtual corrections as well as real gluon emissions to the 
Born--processes.  In addition, gluon--initiated processes also contribute in case of the 
Drell--Yan process as quark--initiated processes in case of Higgs boson production.
The virtual processes are in general UV--divergent due to the 
loop--corrections.
Since one is dealing with light partons inside the incoming hadrons,
the real emission processes suffer from collinear divergences.
In addition, the standard soft divergences appear in both real emission  
as well as virtual contributions to Born processes.   
All these divergences are regularized using $n$--dimensional regularization.
The UV--divergences are removed by standard UV--renormalization.
This introduces the renormalization scale $\mu_R^2$ both through 
$\alpha_s(\mu_R^2)$ and $\ln(Q^2/\mu_R^2)$--terms.  The soft divergences 
cancel when virtual and real 
emission contributions are added.  The remaining collinear divergences
are removed by the mass factorization procedure.
This introduces a new scale called factorization scale $\mu_F^2$.
Finally one ends up with the regular functions  as 
$\ln(x),\ln(1-x)$, polynomials in $x$ and distribution functions such as
$(\ln^m(1-x)/(1-x))_+$ with $m=0,1$;~$\delta(1-x)$.  

Beyond NLO, the computation becomes more complicated.
First of all there are many more processes that contribute to this
order.  For example one finds, two--loop virtual contributions 
to the  Born terms, one loop virtual corrections to the 
NLO contributions, double real emissions to the Born contributions, and 
single real emissions to the NLO processes. In addition
several new type of processes start contributing from NNLO level onwards.
Due to this, the number of integrations involved increases
enormously and new functions emerge.  
The two--loop virtual corrections to the Born process generate only
constants such as $\zeta_2, \zeta_3, \zeta_4$.
But one--loop corrections to NLO processes and real emissions
to Born--, NLO-- and the new NNLO--processes lead to
hypergeometric functions $F_{2,1}$ which upon integration
generate further higher functions.  The hypergeometric functions
usually result from the one--loop box to NLO processes and
also from angular integration of the real emission processes at the NNLO 
level.
Higher functions are also generated when NLO--splitting functions
are convoluted with the Born-- and NLO--cross sections.  Such
convolutions are required in order to perform mass factorization.
The NNLO results are finally expressible in terms of polynomials 
in $x$, logarithms and  Nielsen integrals $S_{n,p}(x)$ \cite{nielsen}, 
both with various 
argument--functions, defined by
\begin{equation}
\label{eq3aa}
S_{n,p}(x)\!\!=\!\!{(-1)^{n+p-1} \over (n-1)!p!}
\int_0^1 {dz \over z}\! \ln^{n-1}(z)\!
\ln^p(1-zx)
\end{equation} 
The  {\sf weight w} of these functions is defined by
{\sf w = p + n}, where any power of a logarithm counts for {\sf w = 1} as the case for
the denominators $1/x,~1/(1+x)$ and $1/(1-x)$. In a product of functions 
the weights of the 
factors add. The standard polylogarithms are related to the Nielsen integrals
by~:
\begin{eqnarray}
\Li_n(x)&=&{d \Li_{n+1}(x) \over d \ln(x)} \equiv S_{n-1,1}(x)
\nonumber\\[2ex]
&=&
{(-1)^{n-1} \over (n-2)\!} \int_0^1 {dz\over z} \ln^{n-2} (z)
\ln(1-zx) \quad {\rm for}\quad n \geq 2~.
\label{eq2}
\end{eqnarray}
One relates the logarithms to the polylogarithms by simple differentiation:
\begin{eqnarray}
{d \Li_2(\pm x)\over d \ln(x)}&=&\Li_1(\pm x) =-\ln(1\mp x)~,
\nonumber\\[2ex]
\Li_0(x)&=&{x \over 1-x}~.
\label{eq3}
\end{eqnarray}
Similarly, one reduces the weight of a Nielsen integral
by differentiating w.r.t. $\ln(x)$~:
\begin{eqnarray}
{d S_{n,p}(x)\over d \ln(x)} = S_{n-1,p}(x)~.
\label{eq4}
\end{eqnarray}
In addition, distributions as
$\delta(1-x)$ and $(\ln^m(1-x)/(1-x))_+$ with $m=0,1,2,3$ contribute. 
These 
functions and their Mellin transforms can be found in Ref.~\cite{JK}.

The hadronic scattering cross sections are obtained 
using Eq.~(\ref{eq1}). One performs the integration over 
$x_1$ and $x_2$ after folding the perturbativly computed 
coefficient functions with the appropriate 
parton distributions.  This may involve further evaluation of
various Nielsen integrals  and increases
the complexity of the numerical evaluation of the hadronic
cross sections. In the next section we will study the structure of
these corrections in  Mellin space using algebraic
identities which relate the resulting finite harmonic sums.
We will present an alternative treatment of the evaluation 
of the total cross sections upto NNLO  by working in Mellin space.  
Such techniques have been used in the past to compute
deep--inelastic scattering cross sections. They are also
found to be most suitable for various resummation programs \cite{JV}.
\footnote{Heavy flavor contributions in Mellin space were treated in 
\cite{JBSA}.}
\section{Mellin Moments}

\vspace{1mm}\noindent
The Mellin--transform \cite{mellin} of a given function 
$F(x)$ is defined by
\begin{equation}
{\Mvec}\big[F\big](N)=\int_0^1 dx x^{N-1} F(x)~.
\end{equation}
The Mellin convolution of two functions $F_1(x),F_2(x)$ is given by
\begin{eqnarray} 
\label{eq8A}
\left[F_1 \otimes F_2\right] (x) = \int_0^1 dx_1 \int_0^1 dx_2 F_1(x_1) 
F_2(x_2)
\delta(x-x_1 x_2)~.
\label{eq5}
\end{eqnarray}
$\Mvec\left[[F_1 \otimes F_2\right](x)](N)$ reduces to 
the product of Mellin moments of $F_1(x)$ and $F_2(x)$, i.e.
\begin{eqnarray}
{\Mvec}\big[F_1 \otimes F_2\big](N) = {\Mvec}\big[F_1\big](N) 
\cdot
{\Mvec}\big[F_2\big](N)~. 
\label{eq6}
\end{eqnarray}
One may generalize this property for the convolution of
$m$ functions~: 
\begin{eqnarray}
\left[F_1 \otimes F_2 \otimes ... \otimes F_m\right] (x) &=& \int_0^1 dx_1 
\int_0^1 
dx_2 ...\int_0^1 
dx_m \int_0^1 dz F_1(x_1) F_2(x_2) ...F_m(x_m) 
\nonumber\\ & &~~~~~~~~~~~~~~~~~~~~~~~~~~~~~~~~~~ \times
\delta(x- x_1 x_2 ...x_m)~,
\label{eq7}
\end{eqnarray}
\begin{eqnarray}
{\Mvec}\big[F_1 \otimes F_2 \otimes ...F_m\big](N) =
\prod_{k=1}^m
{\Mvec}\big[F_k\big](N)~.
\label{eq8}
\end{eqnarray}
Indeed, already the multiple convolution of 
rather simple functions in $x$--space may lead to complicated expressions,
cf. Ref.~\cite{JBHK1}. Contrary to that, the representation (\ref{eq8}) is 
straightforward.

Due to mass--factorization the QCD--improved collinear parton model 
relates the hadronic 
cross sections via Mellin convolutions 
to the partonic cross sections and the parton distribution functions.
Hence the cross section in the Mellin--$N$ space becomes
\begin{eqnarray} 
\label{eq12aa}
{\Mvec}\big[\sigma\big](N,Q^2)&=&{\Mvec}\big[f_a\big](N,\mu^2) {\Mvec}\big[f_b\big](N,\mu^2)
{\Mvec}\big[\Delta_{ab}\big]\Bigg(N,{Q^2 \over \mu^2}\Bigg)~.
\label{eq9}
\end{eqnarray} 
The Mellin moments of these functions
can be analytically continued \cite{JB3} to complex values of $N$ 
so that
one can use various analyticity properties of these functions in
complex $N$--space to evaluate them efficiently.  To retrieve back
the full $x-$dependent result for
(\ref{eq9}), we have to take the inverse Mellin transform 
as a numerical contour integral around all singularities in $N$.

For our analysis, the starting point is Eq.~(\ref{eq8A}) with given
parton densities $f_a(x,\mu^2)$ and known
coefficient functions $\Delta_{ab}(x,Q^2)$ computed upto NNLO in 
perturbative QCD.  
We then compute the Mellin moments of these functions in $N$--space
and analytically continue them to complex $N$--space.  At the end,
we use Eq.~(\ref{eq9}) and 
perform the inverse Mellin transformation to arrive at
the results in $x$--space using a suitable contour in the complex 
$N$--space.
Since only one integral has to be carried out numerically, the evaluation 
can be performed very fast.

Before we study the Mellin moment of the coefficient functions,
we would like to make a few remarks on the parton densities.
As is well known, the parton densities are fitted as functions of $x$
using the available deeply inelastic scattering data.
The parton distributions are determined at some scale $Q_0^2$ along with 
the QCD--scale, $\Lambda_{\rm QCD}$.
The renormalization group equations for mass factorization 
in $x$--space
\begin{eqnarray}
\label{AP1}
\mu^2 {d f_{a/P}(x,\mu^2) \over d \mu^2}
={\alpha_s(\mu^2) \over 4 \pi} 
\sum_{b=q,\bar q,g}
\int_x^1 {dz \over z} 
P_{ab}(z,\alpha_s(\mu^2))
f_{b/P}(z,\mu^2) \quad \quad \quad a=q,\bar q,g
\label{eq10}
\end{eqnarray}
relate the parton densities at $Q_0^2$ to those at the scale $\mu^2$.
The splitting functions
$P_{ab}(z,\alpha_s(\mu^2))$ are computable
order by order in perturbation theory:
\begin{eqnarray}
P_{ab}(z,\alpha_s(\mu^2))=\sum_{n=0}^\infty 
\left({\alpha_s(\mu^2)\over 4 \pi} \right)^n P^{(n)}_{ab}(z)~.
\label{eq11}
\end{eqnarray}
Instead solving the integro--differential equations (\ref{AP1})
one may Mellin--transform these equations to
\begin{eqnarray}
\mu^2
{d {\Mvec}\big[f_{a/P}\big](N,\mu^2)\over d\mu^2}=
{\alpha_s(\mu^2) \over 4 \pi} \sum_{b=q,\bar q,g}
{\Mvec}\big[ P_{ab}\big](N,\alpha_s(\mu^2))
{\Mvec}\big[ f_{b/P}\big](N,\mu^2)~.
\label{eq12}
\end{eqnarray}
The solution of (\ref{eq12}) for ${\Mvec}\big[f_{a/P}\big](N,\mu^2)$
is straightforward as it is now just a first 
order differential equation.  The parton densities in $x-$space
are obtained from the solutions ${\Mvec}\big[f_{a/P}\big](N,\mu^2)$
by an inverse Mellin transformation.  Since we are dealing
with cross sections in $N-$space given in (\ref{eq12aa}), 
the solutions ${\Mvec}\big[f_{a/P}\big](N,\mu^2)$  can be used
for further analysis directly.  

The next task is to compute
the Mellin moments of the known coefficient functions for the different 
hard processes in case of massless fermions to 2--loop order
which were usually computed in $x-$space.

Let us  start with the Drell--Yan process.  We present here
the relevant formulae for both unpolarized as well as polarized
cross sections.  The Drell--Yan process is given by
\begin{eqnarray}
H_1(P_1)+H_2(P_2) \rightarrow l^+(k_1)+l^-(k_2)+X~,
\label{eq13}
\end{eqnarray}
where $H_i$ are the incoming hadrons with momenta $P_i$.
$k_i$, with $i = 1,2$, are the momenta of final state leptons
$l^+l^-$, respectively.  Since the dominant contribution
is through the $\gamma$--exchange $s-$channel processes\footnote{ 
The $Z$--exchange processes can be incorporated in a similar way.}, 
we find the cross section can be expressed as
\begin{eqnarray}
{d \left(\Delta\right) 
\sigma^{DY}(x,Q^2) \over dQ^2} = {4 \pi \alpha^2 \over
3 N_c Q^2 S} \left(\Delta\right) W^{DY}(x,Q^2),
\label{eq14}
\end{eqnarray}
where $\left(\Delta\right)$ denotes the polarized case, 
$Q^2$ is the invariant mass 
of the di--lepton,
\begin{eqnarray}
S=(P_1+P_2)^2, \quad \quad \quad x ={Q^2 \over S}~,
\label{eq15}
\end{eqnarray}
$N_c$  is the number of colors, and $\alpha$ is the electromagnetic 
coupling constant.
The scaling function, usually called the hadronic Drell--Yan structure 
functions
$\left(\Delta\right) W^{DY}(x,Q^2)$, is related to the coefficient 
functions
as
\begin{eqnarray}
\left(\Delta \right) W^{DY}(x,Q^2) &=&\sum_{a,b=q,\bar q,g}
\int_0^1 dx_1 \int_0^1 dx_2 \int_0^1 dz 
\left(\Delta\right) f_{a/H_1}(x_1,\mu^2) 
\left(\Delta\right)f_{b/H_2}(x_2,\mu^2) 
\nonumber\\[2ex]&&
\times 
\left(\Delta\right)
\Delta_{ab}^{DY}\left(z,Q^2,\mu^2\right)
\delta(x-z x_1 x_2)~,
\label{eq16}
\end{eqnarray}
where we have set the renormalization scale to be equal to the 
factorization
scale.  Alternatively, we can express the above equation
as
\begin{eqnarray}
\left(\Delta \right) W^{DY}(x,Q^2) &=&\sum_{a,b=q,\bar q,g}
\int_x^1 {dy \over y} \left(\Delta\right) \Phi_{ab}(y,\mu^2)
\left(\Delta\right)\Delta_{ab}^{DY}\left({x \over y},Q^2,\mu^2\right)~,
\label{eq17}
\end{eqnarray}
where the flux $\Phi_{ab}$ is defined by
\begin{eqnarray}
\left(\Delta\right) \Phi_{ab}(y,\mu^2)=
\int_y^1 {dz \over z}
\left(\Delta\right)f_{a/H_1}(z,\mu^2) 
\left(\Delta\right)f_{b/H_2}\left({y \over z},\mu^2\right)~.
\label{eq18}
\end{eqnarray}

The Mellin moment 
of $x^{-1} \left[d \left(\Delta\right) \sigma^{DY}
(x,Q^2)/dQ^2\right]$ reduces to
\begin{eqnarray}
{\Mvec}\left[{1 \over x} {d \left(\Delta\right)\sigma^{DY}\over d Q^2}\right]
(N,\mu^2)
={4 \pi \alpha^2 \over 3 N_c Q^4} {\Mvec}\left[ \left(\Delta\right) W^{DY}
\right](N,\mu^2)~,
\label{eq19}
\end{eqnarray}
where 
\begin{eqnarray}
\label{eqMM}
{\Mvec}\left[ \left(\Delta\right) W^{DY}\right](N,\mu^2)
={\Mvec}\left[\left(\Delta\right) \Phi_{ab}\right](N,\mu^2) 
{\Mvec}\left[\left(\Delta\right) \Delta_{ab}^{DY}\right](N,\mu^2)~. 
\label{eq20}
\end{eqnarray}
The flux 
${\Mvec}\left[\left(\Delta\right) \Phi_{ab}\right](N,\mu^2)$
can be computed using (\ref{eq12}) with the input parton
densities as boundary conditions.

We now present the total cross section for 
both scalar $(H)$ and pseudo--scalar  $(A)$ Higgs bosons 
at hadron colliders. 
The process is given by
\begin{eqnarray}
H_1(P_1)+H_2(P_2) \rightarrow B + X~,
\label{eq21}
\end{eqnarray}
where $B=H,A$.
The total cross section for Higgs boson production is found to be
\begin{eqnarray}
\sigma_{tot}^{B}(x,m^2)={\pi G_B^2 \over 8 (N^2_c-1)} \sum_{a,b=q,\bar 
q,g}
\int_x^1 dx_1 \int_{x \over x_1}^1 dx_2 f_{a/H_1} (x_1,\mu^2)
f_{b/H_2}(x_2,\mu^2) 
\nonumber\\[2ex]
\times \Delta_{ab,B}\left({x \over x_1 x_2},m^2,\mu^2\right)~,
\quad \quad \quad B=H,A,
\label{eq22}
\end{eqnarray} 
where $m^2$ is the mass of the Higgs boson and the scaling variable
is defined by
\begin{eqnarray} 
x={m^2 \over S} ,\quad \quad \quad S=(P_1+P_2)^2~.
\label{eq23}
\end{eqnarray} 
$\Delta_{ab,B}(x,m^2,\mu^2)$ is the partonic coefficient function.
Again we have set both renormalization and factorization scale to be 
equal. 
The overall constant $G_B$ is
\begin{eqnarray}
G_B=-2^{5/4} a_s\left(\mu_R^2\right) G_F^2 \tau_B F_B(\tau_B){\cal C}_B
\left(a_s(\mu_R^2),\mu_R^2,m_t^2\right)~,
\label{eq24}
\end{eqnarray}
with 
\begin{eqnarray} 
a_s(\mu_R^2)={\alpha_s(\mu_R^2)\over 4 \pi}~,
\label{eq25}
\end{eqnarray}
where $\mu_R$ is the renormalization scale.
$G_F$ is the Fermi constant and the functions $F_B(\tau)$ are given by
\begin{eqnarray}
F_H(\tau)&=&1+(1-\tau) f(\tau),\quad \quad \quad  F_A(\tau)=f(\tau) \cot\beta
\nonumber\\[2ex]
\tau&=& {4 m_t^2 \over m^2}
\nonumber\\[2ex]
f(\tau)&=& \arcsin^2{1 \over \sqrt \tau }, \quad \quad {\rm for} \quad 
\quad 
\tau \geq 1, 
\nonumber\\[2ex]
f(\tau)&=&- {1 \over 4} \left( \ln{1-\sqrt{1-\tau} \over 1+\sqrt{1-\tau}}
+\pi i\right)^2 \quad \quad 
\tau \le 1, 
\label{eq26}
\end{eqnarray}
where $\beta$ is the mixing angle in the two--Higgs doublet model. 
$m_t$ is the mass of the top quark.
The  coefficient ${\cal C}_B$ is given by
\begin{eqnarray}
{\cal C}_H\left(a_s(\mu_R^2) ,m_t^2\right)&=&
1+a_s^{(5)}(\mu_R^2) \Big[5 C_A-3 C_F\Big]+\left(a_s^{(5)}(\mu_R^2)\right)^2
\Bigg[{27 \over 2 } C_F^2 
\nonumber \\[2ex] & &
-{100 \over 3} C_A C_F + {1063 \over 36} C_A^2 
-{4 \over 3 } C_F T_f
-{5 \over 6} C_A T_f +\big( 7 C_A^2 -11 C_A C_F  \big)\ln {\mu_R^2 \over 
m_t^2}
\nonumber \\[2ex]
& &
+n_f T_f \Big(-4 C_F -{47 \over 9 } C_A + 8 C_F \ln{\mu_R^2 \over m_t^2}
\Big)\Bigg],
\nonumber\\[2ex]
{\cal C}_A\left(a_s(\mu_R^2) ,m_t^2\right) &=& 1,
\label{eq27}
\end{eqnarray}
\cite{HMAS1},
where $a_s^{(5)}$ refers to the five--flavor number scheme.
The color factors are
\begin{eqnarray}
C_A=N_c, \quad \quad \quad C_F={N^2_c -1 \over 2}, \quad \quad T_f={1 
\over 2}~. 
\label{eq28}
\end{eqnarray}
Eq.~(\ref{eq22}) can be expressed in a compact form as
\begin{eqnarray}
{1 \over x} \sigma_{tot}^B(x,m^2)=
{\pi G_B^2 \over 8 (N^2-1)} \sum_{a,b=q,\bar q,g}
\int_x^1 {dy \over y} \Phi_{ab}(y,\mu^2)  {y \over x}
\Delta_{ab,B}\left({x \over y},m^2,\mu^2\right)~.
\label{eq29}
\end{eqnarray} 
The Mellin transform of  (\ref{eq29}) becomes
\begin{eqnarray}
\label{eq33aa}
{\Mvec}\left[\sigma^B_{tot} \right](N,m^2)
={\pi G_B^2 \over 8 (N^2-1)} \sum_{ab=q,\bar q,g}
{\Mvec}\left[\Phi_{ab}\right](N+1,\mu^2) 
{\Mvec}\left[\Delta_{ab,B}\right](N,m^2,\mu^2)~.
\label{eq30}
\end{eqnarray}
The flux ${\Mvec}\left[\Phi_{ab}\right]$ can be extracted 
from the solution of the evolution equations (\ref{eq12aa}) and 
${\Mvec}\left[\Delta_{ab,B}\right]$ can be computed analytically
from the known functions $\Delta_{ab,B}(x,m^2,\mu^2)$, see appendix 3,4.

\section{Finite Harmonic Sums}

\vspace{1mm}\noindent
We are dealing with the total scattering cross sections for 
the unpolarized and polarized Drell--Yan process and hadronic 
(pseudo)scalar Higgs boson production in the heavy--mass limit,
which depend only on two variables, $x = Q^2/s$ and $Q^2$, the
invariant mass squared of the final state.
The coefficient functions upto NNLO 
contain a large class of $x$--space functions~\cite{JK}. Their complexity 
reaches nearly $80$, the number of all alternating and non--alternating 
finite 
harmonic sums  \cite{JK,JB1,JB04} of weight {\sf w $\leq$ 4}. The class  
contains simple functions of the form
$(1 \pm x)^{-1},\ln^m(x)$, $\ln^m(1-x)$, and 
$(\ln^m(1-x)/(1-x))_+,
m=0 \ldots 3$. More complicated examples 
are  
weighted Nielsen--integrals 
$S_{n,p}(x)/(1 \mp x)$.
The Mellin transforms of these functions were evaluated in \cite{JK}. In 
the following we present some examples. 

The Mellin moment of the function $1/(1+x)$ can be computed
as follows:
\begin{eqnarray}
\int_0^1 dx x^{N-1} {1 \over (1+x)} &=&\int_0^1 dx x^{N-1} \sum_{i=0}
^\infty (-1)^i x^i
= \sum_{i=0}^\infty {(-1)^{i} \over i+N}
\nonumber\\[2ex]
&=& (-1)^{N-1} \left[\sum_{i=1}^{N-1}{(-1)^i \over i} 
-\sum_{i=1}^\infty{(-1)^i \over i} \right]
=(-1)^{N-1} \left[ S_{-1}(N-1)+ \ln(2)\right]~.
\label{eq31}
\end{eqnarray}
Likewise one obtains
\begin{eqnarray}
\int_0^1 dx x^{N-1} {1 \over (1-x)_+}
&=&\int_0^1 dx \frac{x^{N-1}-1}{1-x}
=\int_0^1 dx (x^{N-1}-1) \sum_{i=0}^\infty x^i
\nonumber\\[2ex]
&=&\sum_{i=0}^\infty \left({1 \over i+N}-{1 \over i+1}\right)
=-S_1(N-1)~,
\label{eq32}
\end{eqnarray}
where the finite harmonic sums are defined as
\begin{eqnarray}
S_{k}(N)=\sum_{i=1}^N {1 \over i^k},
\quad \quad \quad
S_{-k}(N)=\sum_{i=1}^N {(-1)^k \over i^k}~.
\label{eq33}
\end{eqnarray}
Similarly, we compute the Mellin moment of $x^r \ln(1-x)$ by
expanding 
\begin{equation}
\ln(1-x)= - \sum_{i=1}^\infty \frac{x^i}{i}
\end{equation}
using
partial fractions and changing the limits of the sums~:
\begin{eqnarray}
\int_0^1  dx x^{N-1} x^r \ln(1-x)&=&-\sum_{i=1}^\infty {1 \over i 
(i+N+r)}
={1 \over N+r} \sum_{i=1}^\infty \left({1 \over i+N+r}-{1 \over i}\right)
\nonumber\\[2ex]
&=&-{1 \over N+r} \sum_{i=1}^{N+r} {1 \over i}=-{S_1(N+r) \over N+r}~.
\label{eq34}
\end{eqnarray}
One may compute the Mellin moment of the distribution
\begin{eqnarray}
\int_0^1 dx x^{N-1} \left({\ln(1-x)\over 1-x}\right)_+
&=& \int_0^1 dx \left[x^{N-1}-1\right] {\ln(1-x)\over 1-x}
={N-1 \over 2} \int_0^1 x^{N-2} \ln^2(1-x)
\nonumber\\[2ex]
&=&{N-1 \over 2} \sum_{i,j=1}^\infty {1 \over i j (i+j+N-1)}~.
\label{eq35}
\end{eqnarray}
The summation is performed after partial fractioning
and shifting the summation limits and one arrives at
\begin{eqnarray}
\int_0^1 dx x^{N-1} \left({\ln(1-x)\over 1-x}\right)_+
&=& S_{1,1}(N-1)~.
\label{eq36}
\end{eqnarray}
The latter sum is a multiple (nested)
harmonic sum defined by
\begin{eqnarray}
S_{m,m_1,m_2,...,m_k}(N)=\sum_{i=1}^N {S_{m_1,m_2,...,m_k}(i) \over 
i^m}~,~\forall m_i > 0~.
\label{eq37}
\end{eqnarray}
In general, also alternating nested harmonic sums contribute, which are 
labeled with as well negative indices $k_i < 0$, 
\begin{eqnarray}
S_{k_1...k_m}(N)&=&\sum_{n_1=1}^N {[\si(k_1)]^{n_1} \over n_1^{|k_1|}}
                 \sum_{n_2=1}^{n_1} {[\si(k_2)]^{n_2} \over n_2^{|k_2|}}
\cdot \cdot \cdot\sum_{n_m=1}^{n_m-1} {[\si(k_m)]^{n_m} \over 
n_m^{|k_m|}}~,
\label{eq38}
\end{eqnarray}
with $k_l, l \not=0$.  Similarly one computes
Mellin moments of more complicated function.  For example,
\begin{eqnarray}
{\Mvec}\left[\left({\ln^3(1-x)\over 1-x}\right)_+\right](N)&=&
{1\over 4} S_1^4(N-1)
        +{3\over 2} S_1^2(N-1) S_2(N-1)
        +{3\over 4} S_2^2(N-1)
\nonumber\\[2ex]
&&        +2 S_1(N-1) S_3(N-1)
        +{3\over 2} S_4(N-1)~,
\label{eq39}
\end{eqnarray}
where the right hand side contains only single finite harmonic sums.
The Mellin moments of Nielsen integrals can be mostly done
by relating them to simpler moments after an integration by
parts:
\begin{eqnarray}
\int_0^1 dx x^{N-1} S_{1,2}(-x)={1 \over N}
\frac{\zeta_3}{8}-{1 \over 2 N}
{\Mvec}\Big[\ln^2(1+x)\Big](N)
\end{eqnarray}
The Mellin transforms of individual functions in $x$--space may contain 
complicated sums~\cite{JK}, which cancel in combinations. As an example we 
mention 
\begin{eqnarray}
&&\int_0^1 dx x^{N-1} \Bigg[ S_{1,2}(-x) +  \Li_2(-x) \ln(1+x) 
+ \frac{1}{2} \ln(x) 
\ln^2(1+x)\Bigg]\nonumber\\
&&= \frac{(-1)^{N-1}}{N} \Bigg\{
 S_{-1,2}(N) + \frac{\zeta_2}{2} \left[S_1(N) - S_{-1}(N)\right]\Biggr\}
 + \frac{1+(-1)^{N-1}}{N}\left[\frac{\zeta_3}{8} - \frac{\ln(2) \zeta_2}{2}\right] 
\label{eq40}
\end{eqnarray}
in which the sum $S_{-1,1}(N)$ does not occur unlike the case for 
$\Mvec[\Li_2(-x) \ln(1+x)](N)$.
We finally mention recursive integral--representations for the finite 
harmonic sums and weighted power sums, cf. \cite{JK},
\begin{eqnarray} 
S_{\pm k}(N)&=\!\!&\!\!\int_0^1{dx_1\over x_1} \cdot \cdot \cdot \int_0^{x_{k-1}}
{(\pm x_k)^N-1\over x_k \mp 1}
\nonumber\\[2ex]
&\!\!\!=\!\!\!&{(-1)^{k-1} \over (k-1)!}\int_0^1 dx \ln^{k-1}(x)
{(\pm x)^N-1 \over x \mp 1}~,
\label{eq41}\\
\sum_{k=1}^N{(\pm x)^k \over k^l}&\!\!\!=\!\!\!&
{(-1)^{l-1} \over (l-1)!}\!\int_0^x \!dz \!\ln^{l-1}(z)
{(\pm z)^N-1 \over z \mp 1}~.
\label{eq42}
\end{eqnarray} 
The translation of the 2--loop Wilson coefficients \cite{DY,HGS} from 
$x$-- to $N$--space is performed using the tables given in \cite{JK} for 
the individual functions. One observes the cancellation of a series of 
sums. In particular we would like to mention, that multiple sums with the 
index 
$\{-1\}$  do not occur in the final results up to {\sf w=4}.

\section{Algebraic Relations}

\vspace{1mm}\noindent
Finite harmonic sums obey algebraic equations, see e.g. \cite{JB1}. 
The more simple relations for finite harmonic sums can be 
found in \cite{euler,NIELSHB,GIRG,JK}. Finite harmonic sums obey a shuffle 
algebra. To further simplify the expressions for the Wilson coefficients 
dealt with in the present paper the following relations are used for sums 
with two
\begin{eqnarray}
 S_{1,1}(N)&=&{1 \over 2}(S_{1}(N)^2+S_{2}(N))
\nonumber\\[2ex]
 S_{-1,1}(N)&=&-S_{1,-1}(N)+S_{1}(N)S_{-1}(N)
+S_{-2}(N)
\nonumber\\[2ex]
 S_{-1,-2}(N)&=&-S_{-2,-1}(N)+S_{-2}(N)S_{-1}(N)
+S_{3}(N)
\nonumber\\[2ex]
 S_{1,2}(N)&=&-S_{2,1}(N)+S_{2}(N)S_{1}(N)
+S_{3}(N)
\nonumber\\[2ex]
 S_{-1,2}(N)&=&-S_{2,-1}(N)+S_{2}(N)S_{-1}(N)
+S_{-3}(N)
\nonumber\\[2ex]
 S_{1,-2}(N)&=&-S_{-2,1}(N)+S_{-2}(N)S_{1}(N)
+S_{-3}(N)
\nonumber\\[2ex]
 S_{2,2}(N)&=&{1 \over 2}(S_{2}(N)^2+S_{4}(N))
\nonumber\\[2ex]
 S_{-2,-2}(N)&=&{1 \over 2}(S_{-2}(N)^2+S_{4}(N))
\nonumber\\[2ex]
 S_{2,-2}(N)&=&-S_{-2,2}(N)+S_{-2}(N)S_{2}(N)
+S_{-4}(N)
\nonumber\\[2ex]
 S_{-1,-3}(N)&=&-S_{-3,-1}(N)+S_{-3}(N)S_{-1}(N)
+S_{4}(N)
\nonumber\\[2ex]
 S_{1,3}(N)&=&-S_{3,1}(N)+S_{3}(N)S_{1}(N)
+S_{4}(N)
\nonumber\\[2ex]
 S_{-1,3}(N)&=&-S_{3,-1}(N)+S_{3}(N)S_{-1}(N)
+S_{-4}(N)
\nonumber\\[2ex]
 S_{1,-3}(N)&=&-S_{-3,1}(N)+S_{-3}(N)S_{1}(N)
+S_{-4}(N)
\label{eq44}
\end{eqnarray}
and three indices~:
\begin{eqnarray}
 S_{1,2,1}(N)&=&-2S_{2,1,1}(N)+S_{3,1}(N)
       +S_{1}(N)S_{2,1}(N)+S_{2,2}(N)
\nonumber\\[2ex]
 S_{1,1,2}(N)&=&S_{2,1,1}(N)
           +\left[{1 \over 2}(S_{1}(N)(S_{1,2}(N)
-S_{2,1}(N))+S_{1,3}(N)-S_{3,1}(N)\right]
\nonumber\\[2ex]
 S_{1,-2,1}(N)&=&-2S_{-2,1,1}(N)+S_{-3,1}(N)
     +S_{1}(N)S_{-2,1}(N)+S_{-2,2}(N)
\nonumber\\[2ex]
 S_{1,1,-2}(N)&=&S_{-2,1,1}(N)+S_{-2}(N)S_{2}(N)
    -S_{-2,2}(N)-S_{-2}(N)S_{1,1}(N)
\nonumber\\[2ex]&&
            +S_{1}(N)S_{1,-2}(N)+S_{1,-3}(N)
-S_{1}(N)S_{-3}(N)~.
\label{eq46}
\end{eqnarray}
At  intermediate stages of
the computation we encounter various more complicated sums
such as $S_{1,-1,2}$, $S_{-1,-1,-2}$, $S_{-1,-2,-1}$, $S_{-2,-1,-1}$,
$S_{2,-1,1}$, $S_{1,2,-1}$, $S_{2,1-1}$, $S_{-1,1,2}$, $S_{-1,1,2}$,
$S_{-1,-3}$, and $S_{-1,3}$.  
All these sums do finally occur in symmetric combinations, which are 
{polynomials} of {\sf single} harmonic sums.
At the level of 3--fold sums only $S_{-2,1,1}$ and $S_{2,1,1}$ remain.

\section{Coefficient Functions in \boldmath{$N$}-Space}

\vspace{1mm}\noindent
Let us now express the Wilson coefficients in terms of the remaining 
harmonic sums in terms of polynomials and as rational functions in $N$.
Furthermore, we list the respective harmonic sums in terms of their 
Mellin--transforms, cf. \cite{JK}. 

The single harmonic sums are~:
\begin{eqnarray}
S_{-4}(N)&=&(-1)^{N+1}{1 \over 6} 
{\Mvec}\left[{\ln^3(x)\over 1+x}\right](N+1)
-{7 \over 20} \zeta_2^2
\nonumber\\[2ex]
S_{-3}(N)&=&(-1)^N{1 \over 2} 
{\Mvec}\left[{\ln^2(x)\over 1+x}\right](N+1)
-{3 \over 4} \zeta_3
\nonumber\\[2ex]
S_{-2}(N)&=&(-1)^{N+1}
{\Mvec}\left[{\ln(x)\over 1+x}\right](N+1)
-{1 \over 2} \zeta_2
\nonumber\\[2ex]
S_{-1}(N)&=&(-1)^N 
{\Mvec}\left[{1\over 1+x}\right](N+1)
-\ln(2)
\nonumber\\[2ex]
S_{4}(N)&=&{1 \over 6}
{\Mvec}\left[{\ln^3(x)\over 1-x}\right](N+1)
+{2 \over 5} \zeta_2^2
\nonumber\\[2ex]
S_{3}(N)&=&-{1 \over 2}
{\Mvec}\left[{\ln^2(x)\over 1-x}\right](N+1)
+\zeta_3
\nonumber\\[2ex]
S_{2}(N)&=&
{\Mvec}\left[{\ln(x)\over 1-x}\right](N+1)
+\zeta_2
\nonumber\\[2ex]
S_{1}(N)&=&-
{\Mvec}\left[\left({1\over 1-x}\right)_+\right](N+1)~.
\label{eq47}
\end{eqnarray}
These harmonic sums can be solely expressed in terms of  Euler's
$\psi$--function and the $\beta$--function~\cite{NIELSHB} and their
derivatives, which is related to the former combining two
$\psi$--functions with shifted argument.
These functions represent at the same time the analytic continuation of
these harmonic sums~:
\begin{eqnarray}
S_{k}(N)  &=& \frac{(-1)^{k+1}}{(k-1)!} \psi^{(k-1)}(N+1) + \zeta(k) \\
S_{-1}(N) &=& (-1)^N \beta(N+1) - \ln(2) \\
S_{-k}(N) &=& \frac{(-1)^{N+k-1}}{(k-1)!} \beta^{(k-1)}(N+1) - 
\left(1-\frac{1}{2^{k-1}}\right) \zeta_k,~~k \geq 2~,
\\
\beta(z) &=& \frac{1}{2} \left[ \psi\left(\frac{1+z}{2}\right) - 
\psi\left(\frac{z}{2}\right)\right]~.
\end{eqnarray}
The following five double sums occur~:
\begin{eqnarray}
S_{-3,1}(N)&=&(-1)^N 
{\Mvec}\left[{\Li_3(x)\over 1+x}\right](N+1)
+\zeta_2 S_{-2}(N)
-\zeta_3 S_{-1}(N)
-{3 \over 5} \zeta_2^2
+2 \Li_4\left({1 \over 2}\right)
\nonumber\\[2ex]&&
+{3 \over 4} \zeta_3 \ln(2)
-{1 \over 2} \zeta_2 \ln^2(2)
+{1 \over 12} \ln^4(2)
\nonumber\\[2ex]
S_{-2,1}(N)&=&-(-1)^N 
{\Mvec}\left[{\Li_2(x)\over 1+x}\right](N+1)
+\zeta_2 S_{-1}(N)
-{5 \over 8} \zeta_3
+ \zeta_2 \ln(2)
\nonumber\\[2ex]
S_{-2,2}(N)&=&-(-1)^N 
{\Mvec}\Bigg[{1\over 1+x}\Big(2 \Li_3(x)
-\ln(x) \Big(\Li_2(x) 
+\zeta_2\Big)\Big)\Bigg](N+1)
+\zeta_2 S_{-2}(N)
\nonumber\\[2ex]&&
+2 \zeta_3 S_{-1}(N)
+{71 \over 40} \zeta_2^2
-4 \Li_4\left({1 \over 2}\right)
-{3 \over 2} \zeta_3 \ln(2)
+\zeta_2 \ln^2(2)
-{1 \over 6} \ln^4(2)
\nonumber\\[2ex]
S_{2,1}(N)&=& 
{\Mvec}\left[\left({\Li_2(x)\over 1-x}\right)_+\right](N+1)
+\zeta_2 S_{1}(N)
\nonumber\\[2ex]
S_{3,1}(N)&=& -{1 \over 2}
{\Mvec}\left[{\Li_2(x)\ln(x)\over 1-x}\right](N+1)
+\zeta_2 S_{2}(N)
-{1 \over 4} S_2^2(N)
-{1 \over 4} S_4(N)
-{3 \over 20} \zeta_2^2
\label{eq48}
\end{eqnarray}
Two triple sums contribute~:
\begin{eqnarray}
S_{-2,1,1}(N)&=&-(-1)^N 
{\Mvec}\left[{S_{1,2}(x)\over 1+x}\right](N+1)
+\zeta_3 S_{-1}(N)
-\Li_4\left({1 \over 2}\right)
+{1 \over 8} \zeta_2^2
+{1 \over 8} \zeta_3 \ln(2)
\nonumber\\[2ex] &&   
+{1 \over 4} \zeta_2 \ln^2(2)
-{1 \over 24} \ln^4(2)
\nonumber\\[2ex]    
S_{2,1,1}(N)&=& 
{\Mvec}\left[\left({S_{1,2}(x)\over 1-x}\right)_+\right](N+1)
+\zeta_3 S_1(N)~.
\label{eq49}
\end{eqnarray}
For the analytic continuation it is 
sufficient to determine $\Mvec[f(x)](N)$ for complex values of $N$ since
\begin{equation}
\frac{\partial^k}{\partial N^k} \Mvec[f(x)](N) = \Mvec\left[\ln^k(x) 
f(x)\right](N)
\end{equation}
is easily obtained analytically. Therefore we will not count the 
associated derivatives as genuinely new functions.
Besides Euler's $\psi$--function the Mellin transforms of five further 
{\sf basic} functions, 
\begin{eqnarray}
{\Li_2(x)  \over 1-x}, \quad  \quad
{\Li_2(x)  \over 1+x}, \quad  \quad
{S_{1,2}(x) \over 1-x}, \quad  \quad
{S_{1,2}(x) \over 1+x},  \quad  \quad
{\Li_3(x) \over 1+x},
\end{eqnarray}
are sufficient to express the different Wilson coefficients dealt with in 
the present paper. The Mellin transforms of these functions were 
calculated in
\cite{JB3} and are denoted by 
$A_k(N)$ with $k=18, 3, 21, 8, 18, 6$. In the appendices we furthermore 
refer to the functions 
$A_5(N)$ and $A_{22}(N)$, which are given by
\begin{eqnarray}
A_5(N) &=& \frac{\partial}{\partial N} A_3(N)~,\\
A_{22}(N) &=& \frac{\partial}{\partial N} A_{18}(N)~.
\end{eqnarray}
The analytic continuation to $N~\epsilon~{\bf C}$ is 
unique~\cite{carlson}.
The hadronic cross sections are obtained calculating the inverse Mellin 
transform of (\ref{eqMM},\ref{eq33aa}) by a numerical contour integral
\begin{eqnarray}
F(x)={1 \over 2 \pi i} 
\int_{c-i\infty}^{c+i \infty} dN x^{-N} \Mvec[F(x)](N)~.
\label{eq51}
\end{eqnarray}
Here the parameter $c$ is the intersection of the contour and the real axis
and is chosen right to the rightmost singularity of the function 
$\Mvec[F(x)](N)$.
The shape of the contour can be deformed at our convenience covering all 
singularities of $\Mvec[F(x)](N)$.  

\section{Conclusion}

\vspace{1mm}\noindent
We have systematically analyzed the mathematical structure
behind the NNLO coefficient functions for the unpolarized and polarized 
Drell--Yan process and hadronic scalar and pseudo--scalar Higgs boson  
production
using Mellin moment techniques.  Use of various algebraic and 
structural identities,
which relate the finite harmonic sums, reduces the complexity of the 
results from 
around 80 functions to only five basic functions, the $\psi$--function and a few 
derivatives thereof. This is very useful both for the understanding of the nature of 
higher order corrections in the massless case and yields expressions which allow to 
perform fast numerical calculations
at high precision for phenomenological applications and fits to data.
The same structures are found in the case of polarized and
unpolarized 2--loop fragmentation functions~\cite{JBVR2}. Together with
the results of \cite{JBSM1} for the Wilson coefficients for unpolarized and polarized 
deeply inelastic scattering it is now shown that these structures are in
common for all known massless 2--loop Wilson coefficients.

\vspace{3mm}\noindent
{\bf Acknowledgment.}\\
We thank S. Moch for discussions. V.R. would like 
to
thank DESY for their kind hospitality extended to him.  
This paper was supported in part by DFG Sonderforschungsbereich
Transregio 9, Computergest\"utzte Theoretische Physik, and EU grant
HPRN--CT--2000--00149.

\include{APPENDIX} 
\newpage 
 
\end{document}

%% file: APPENDIX.tex
\myappendix{\hspace*{-6mm}\large \bf A.1 Unpolarized Drell-Yan 
Coefficient Functions}

\vspace{2mm}\noindent
In this appendix we present the Mellin transforms of the 
coefficient functions
for the unpolarized Drell--Yan Process.   We used the same notation 
as in  
Refs.~\cite{DY}  and refer either to single harmonic sums $S_{k}(N)$ 
or special functions, $A_l(N)$. To avoid the occurrence of boundary terms
in case of $S_{-k}(N)$ we use the functions $\beta^{(k-1)}(N),~~k \geq 1$, 
instead, with $N~\epsilon~{\bf C}$, outside the respective single poles of 
the functions below.
\begin{eqnarray}
\Delta_{ab}=\Delta_{ab}^{(0)}+a_s \Delta_{ab}^{(1)}+a_s^2 \Delta_{ab}^{(2)}
\end{eqnarray}


\newpage
{\hspace*{-6mm}\large \bf A.2 Polarized Drell-Yan Coefficient Functions}

\vspace{2mm}\noindent
In this appendix we present the Mellin transforms of the 
coefficient functions for
of the polarized Drell--Yan process.   We used the same notation 
as in Refs.~\cite{DY}.
\begin{eqnarray}
\delta \Delta_{ab}=\delta \Delta_{ab}^{(0)}
+a_s \delta \Delta_{ab}^{(1)}+a_s^2 \delta \Delta_{ab}^{(2)}
\end{eqnarray}

\begin{eqnarray}
&& \delta \Delta_{q\overline q}^{(0)}=
- \Delta_{q\overline q}^{(0)}
\\[2ex]
&&\delta \Delta_{q\overline q}^{(1)}=
- \Delta_{q\overline q}^{(1)}
\\[2ex]
&&\delta \Delta_{q\overline q}^{(2),S+V}=
- \Delta_{q\overline q}^{(2),S+V}
\\[2ex]
&&\delta \Delta_{q\overline q}^{(2),C_A}=
- \Delta_{q\overline q}^{(2),C_A}
\\[2ex]
&&\delta \Delta_{q\overline q}^{(2),C_F}=
- \Delta_{q\overline q}^{(2),C_F}
\\[2ex]
&&\delta \Delta_{q\overline q,A \overline A}^{(2)}=
- \Delta_{q\overline q,A \overline A}^{(2)}
\\[2ex]
&&\delta \Delta_{q\overline q,A \overline C}^{(2)}=
 \delta \Delta_{q\overline q,A \overline D}^{(2)}
=-\Delta_{q\overline q,A \overline D}^{(2)}
\\[2ex]
&&\delta \Delta_{q\overline q,B \overline B}^{(2)}=
=-\Delta_{q\overline q,B \overline B}^{(2)}
\\[2ex]
&&\delta \Delta_{q\overline q,B \overline C}^{(2)}=
 \delta \Delta_{q\overline q,B \overline D}^{(2)}
=-\Delta_{q\overline q,B \overline C}^{(2)}
\\[2ex]
&&\delta \Delta_{q g}^{(2),C_A}=
 C_A T_f \Bigg[ \,{144 \over N^2}
-\,{144 \over N}
-\,{140 \over (N+1)^2}
+\,{142 \over (N+1)}
+\,{6 \over (N+2)}
\nonumber\\[2ex]&&
-\,{94 \over N^3}
+\,{184 \over (N+1)^3}
+\,{44 \over N^4}
+\,{64 \over (N+1)^4}
+\,{24 \over (N+2)^3}
+44 \,{\frac {{ \zeta_2}}{{N}^{2}}}
-96\,{\frac {{ \zeta_2}}{N}}
\nonumber\\[2ex]&&
-18\,{\frac {{ \zeta_3 }}{N}}
+32\,{\frac {{ \zeta_2}}{{(N+1)}^{2}}}
+68\,{\frac {{ \zeta_2}}{(N+1)}}
+36\,{ \frac {{ \zeta_3}}{(N+1)}}
+24\,{\frac {{ \zeta_2}}{(N+2)}}
-36\,{\frac {{ S_1} ( N+2 ) }{{(N+2)}^{2}}}
\nonumber\\[2ex]&&
-{16 \over 3}\,{\frac {{ S_3} ( N+1 ) }{(N+1)}}
+{8 \over 3}\,{\frac {{ S_3} ( N ) }{N}}
-8\,{\frac {{ \beta^{(1)}} ( N+1) }{{N}^{2}}}
+24\,{\frac {{ \beta^{(1)}} ( N+1 ) }{N}}
-16\,{\frac {{ \beta^{(1)}} ( N+2 ) }{{(N+1)}^{2}}}
\nonumber\\[2ex]&&
+24\,{\frac {{ \beta^{(1)}} ( N+2) }{(N+1)}}
+4\,{\frac {{ \beta^{(2)}} ( N+1 ) }{N}}
+8\,{\frac {{ \beta^{(2)}} ( N+2 ) }{(N+1)}}
+12\,{\frac {{ S_1^2} ( N+2 ) }{(N+2)} }
+ \Bigg( -\,{48 \over (N+1)^3}
\nonumber\\[2ex]&&
-\,{176 \over (N+1)^2}
-52\,{\frac {{ S_2} ( N+1) }{(N+1)}}
+32\,{\frac {{ \beta^{(1)}} ( N+2 ) }{(N+1)}}
+\,{142 \over (N+1)}
\nonumber\\[2ex]&&
+56\,{\frac {{ \zeta_2}}{(N+1)}} \Bigg) { S_1} ( N+1 ) 
-24\,{ \frac {{ A_{18}} ( N+1 ) }{(N+1)}}
+12\,{\frac {{ A_{18}} ( N) }{N}}
+ \Bigg( -\,{136 \over N}
-28\,{\frac {{ \zeta_2}}{N}}
-\,{56 \over N^3}
\nonumber\\[2ex]&&
+26\, {\frac {{ S_2} ( N ) }{N}}
+16\,{\frac {{ \beta^{(1)}} ( N+1 ) }{ N}}
+\,{108 \over N^2} \Bigg) { S_1} ( N ) 
+ \Bigg( \,{120 \over N}
-\,{44 \over N^2} \Bigg) { S_2} ( N ) 
+ \Bigg( \,{8 \over (N+1)^2}
\nonumber\\[2ex]&&
+\,{92 \over (N+1)}\Bigg) { S_1^2} ( N+1 ) 
+ \Bigg( -\,{104 \over N}
+\,{44 \over N^2} \Bigg) { S_1^2} ( N ) 
+ \Bigg( -4\,{\frac {{ S_1} ( N ) }{N}}
+{ 8 \over N^2}
-\,{24 \over N}
\nonumber\\[2ex]&&
+\,{24 \over (N+1)}
+\,{8 \over (N+1)^2}
+8\,{\frac {{ S_1} ( N+1) }{(N+1)}} \Bigg) \ln^2\left({Q^2 \over \mu^2}\right )
+ \Bigg(  ( -\,{16 \over (N+1)^2}
\nonumber\\[2ex]&&
-\,{88 \over (N+1)} ) { S_1} ( N+1 ) 
+ \left( \,{100 \over N}
-\,{40 \over N^2}\right) { S_1} ( N ) 
-24\,{\frac {{ S_1^2} ( N+1 ) }{(N+1) }}
+\,{88 \over (N+1)^2}
\nonumber\\[2ex]&&
+8\,{\frac {{ S_2} ( N+1 ) }{(N+1)}}
+\,{24 \over N^3}
+4 \,{\frac {{ \zeta_2}}{N}}
-\,{72 \over (N+1)}
-4\,{\frac {{ S_2} ( N ) }{N }}
-12\,{\frac {{ S_1} ( N+2 ) }{(N+2)}}
+\,{12 \over (N+2)^2}
\nonumber\\[2ex]&&
-\,{52 \over N^2}
+ 12\,{\frac {{ S_1^2} ( N ) }{N}}
-8\,{\frac {{ \beta^{(1)}} ( N+1) }{N}}
+\,{68 \over N}
-8\,{\frac {{ \zeta_2}}{(N+1)}}
+\,{32 \over (N+1)^3}
\nonumber\\[2ex]&&
-16\,{ \frac {{ \beta^{(1)}} ( N+2 ) }{(N+1)}} \Bigg) { \ln\left({Q^2 \over \mu^2}\right )}
+ \Bigg( -\,{116 \over (N+1)}
-\,{32 \over (N+1)^2} \Bigg) { S_2} ( N+1 ) 
-\,{ \frac {{ 26 ~S_1^3} ( N ) }{3 N}}
\nonumber\\[2ex]&&
+\,{\frac {{ 52 ~S_1^3} ( N+ 1 ) }{3 (N+1)}} \Bigg]
\\[2ex] &&
\delta \Delta_{q g}^{(2),C_F}=
 C_F T_f \Bigg[ \,{175 \over N^2}
-\,{627 \over 2 N}
-\,{79 \over (N+1)^2}+\,{331 \over (N+1)}
- \,{60 \over (N+2)^2}
\nonumber\\[2ex]&&
-\,{15 \over 2 (N+2)}
+ \Bigg( \,{104 \over (N+1)}
-\,{84 \over (N+1)^2}\Bigg) { S_1^2} ( N+1 ) 
-\,{81 \over N^3}
-\,{40 \over (N+1)^3}
\nonumber\\[2ex]&&
+\,{34 \over N^4}
- \,{68 \over (N+1)^4}
+ \Bigg( -56\,{\frac {{ \zeta_2}}{N}}
+2\,{\frac {{ S_2} ( N) }{N}}
-\,{48 \over N^3}
+\,{140 \over N^2}
-\,{256 \over N} \Bigg) { S_1} ( N) 
\nonumber\\[2ex]&&
+\,{12 \over (N+2)^3}
+ \Bigg( \,{42 \over N^2}
-\,{110 \over N} \Bigg) { S_1^2} ( N ) 
+28\,{\frac {{ \zeta_2}}{{N}^{2}}}
-126\,{\frac {{ \zeta_2}}{N}}
+ 32\,{\frac {{ \zeta_3}}{N}}
-56\,{\frac {{ \zeta_2}}{{(N+1)}^{2}}}
\nonumber\\[2ex]&&
+128\,{\frac {{ \zeta_2}}{(N+1)}}
-64\,{\frac {{ \zeta_3}}{(N+1)}}
+4\,{\frac {{ S_2} ( N+1) }{{(N+1)}^{2}}}
-6\,{\frac {{ S_2} ( N+2 ) }{(N+2)}}
+{16 \over 3}\,{ \frac {{ S_3} ( N+1 ) }{(N+1)}}
\nonumber\\[2ex]&&
-{8 \over 3}\,{\frac {{ S_3} ( N ) }{N}}
+16\,{\frac {{ \beta^{(1)}} ( N+1 ) }{{N}^{2}}}
-32\,{\frac {{ \beta^{(1)}} ( N+1 ) }{N}}
-32\,{\frac {{ \beta^{(1)}} ( N+2 ) }{{(N+1)}^{2}}}
- 32\,{\frac {{ \beta^{(1)}} ( N+2 ) }{(N+1)}}
\nonumber\\[2ex]&&
+18\,{\frac {{ S_1^2} ( N+2) }{(N+2)}}
+32\,{\frac {{ A_{18}} ( N+1 ) }{(N+1)}}
-16\,{\frac { { A_{18}} ( N ) }{N}}
+ \Bigg( -4\,{\frac {{ S_2} ( N+1 ) } {(N+1)}}
+\,{154 \over (N+1)}
\nonumber\\[2ex]&&
+112\,{\frac {{ \zeta_2}}{(N+1)}}
-\,{48 \over (N+1)^2}+\,{96 \over (N+1)^3} \Bigg) { S_1} ( N+1 ) 
+ \Bigg(  \left( \,{80 \over (N+1)^2}
\right.\nonumber\\[2ex]&&
\left. - \,{72 \over (N+1)} \right) { S_1} ( N+1 ) 
+ \left( \,{96 \over N}
-\,{40 \over N^2} \right) { S_1} ( N ) 
+\,{16 \over N^3}
+36\,{\frac {{ S_1^2} ( N) }{N}}
\nonumber\\[2ex]&&
-72\,{\frac {{ S_1^2} ( N+1 ) }{(N+1)}}
+32\,{\frac {{ \zeta_2}}{N}}
-\,{76 \over (N+1)}
-\,{32 \over (N+1)^3}
-\,{58 \over N^2}
-4\,{\frac {{ S_2} ( N ) }{N}}
\nonumber\\[2ex]&&
+8\,{\frac {{ S_2} ( N+1 ) }{(N+1)}}
-12\,{\frac {{ S_1} ( N+2 ) }{(N+2)}}
+\,{98 \over N}
+\,{12 \over (N+2)^2}
+\,{16 \over (N+1)^2}
\nonumber\\[2ex]&&
-64\,{\frac {{ \zeta_2}}{(N+1)}} \Bigg) { \ln\left({Q^2 \over \mu^2}\right )}
+ \Bigg( \,{4 \over N}-\,{2 \over N^2}\Bigg) { S_2} ( N ) 
+ \Bigg( \,{60 \over (N+2)}
\nonumber\\[2ex]&&
-\,{24 \over (N+2)^2}\Bigg) { S_1} ( N+2 ) 
+ \Bigg( \,{6 \over N^2}
-12\,{\frac {{ S_1} ( N ) }{N}}
-\,{9 \over N}
-\,{12 \over (N+1)^2}
\nonumber\\[2ex]&&
+24\,{\frac {{ S_1} ( N+1) }{(N+1)}} \Bigg) \ln^2\left({Q^2 \over \mu^2}\right )
-\,{\frac {{ 70 ~S_1^3} ( N ) }{3 N}}
+\,{\frac {{ 140 ~S_1^3} ( N+1 ) }{3 (N+ 1)}} \Bigg]
\\[2ex] &&
\delta \Delta_{q \bar q,C\overline C}^{(2)}=
\delta \Delta_{q \bar q,D\overline D}^{(2)}=
\delta \Delta_{q q,C\overline C}^{(2)}=
\delta \Delta_{q q,D\overline D}^{(2)}=
\delta \Delta_{\bar q \bar q,C\overline C}^{(2)}=
\delta \Delta_{\bar q \bar q,D\overline D}^{(2)}
\nonumber\\[2ex]&&
= C_F T_f \Bigg[   \Bigg( \,{10 \over (N+1)}
-\,{10 \over N}
+\,{4 \over N^2}
+\,{4 \over (N+1)^2} \Bigg) \ln^2\left({Q^2 \over \mu^2}\right )
+ \Bigg(  \left( -\,{40 \over (N+1)}
\right.\nonumber\\[2ex]&& \left.
-\,{16 \over (N+1)^2} \right) { S_1} ( N+1 ) 
+ \left( -\,{16 \over N^2}
+\,{40 \over N} \right) { S_1} ( N) 
+\,{16 \over (N+1)^3}
-\,{40 \over (N+1)}
\nonumber\\[2ex]&&
+\,{40 \over N}
-\,{28 \over N^2}
+\,{32 \over (N+1)^2}
+\,{16 \over N^3} \Bigg) { \ln\left({Q^2 \over \mu^2}\right )}
+ \Bigg( -\,{32 \over (N+1)^3}
-\,{64 \over (N+1)^2}
\nonumber\\[2ex]&&
+\,{80 \over (N+1)} \Bigg) { S_1} ( N+1 ) 
+ \Bigg( -\,{80 \over N}
+\,{56 \over N^2}
- \,{32 \over N^3} \Bigg) { S_1} ( N ) 
+ \Bigg( \,{16 \over (N+1)^2}
\nonumber\\[2ex]&&
+40\,\frac{1}{(N+1)} \Bigg) { S_1^2} ( N+1 ) 
+ \Bigg( \,{16 \over N^2}
-\,{40 \over N} \Bigg) { S_1^2} ( N ) 
+ \Bigg( -\,{24 \over (N+1)^2}
\nonumber\\[2ex]&&
-\,{52 \over (N+1)} \Bigg) { S_2} ( N+1 ) 
+ \Bigg( \,{52 \over N}
-\,{24 \over N^2} \Bigg) { S_2} ( N) 
+52\,{\frac {{ \zeta_2}}{(N+1)}}
+24\,{\frac {{ \zeta_2}}{{(N+1)}^{2}}}
\nonumber\\[2ex]&&
+ \,{84 \over N^2}
-\,{131 \over N}
-\,{50 \over N^3}
+\,{36 \over N^4}
-52\,{\frac {{ \zeta_2}}{N}}
+\, {78 \over (N+1)^3}
-\,{34 \over (N+1)^2}
\nonumber\\[2ex]&&
+\,{36 \over (N+1)^4}
+\,{128 \over (N+1)}
+24\,{\frac {{ \zeta_2}}{{N}^{2}}}
+\,{3 \over (N+2)} \Bigg ]
\\[2ex] &&
\delta \Delta_{q \bar q,C\overline D}^{(2)}=
-\delta \Delta_{q q,C\overline D}^{(2)}=
-\delta \Delta_{\bar q \bar q,C\overline D}^{(2)}
\\[2ex] &&
= C_F T_f \Bigg[ -\,{16 \over N^2}
+\,{32 \over N}
-\,{32 \over (N+1)}
+\,{8 \over (N+1)^3}
-\,{32 \over (N+1)^4} 
\nonumber\\[2ex]&&
+192\,{\frac {{ \beta} ( N+1 ) { \zeta_2}}{N}}
+16\,{\frac {{ \zeta_2}} {{N}^{2}}}
-8\,{\frac {{ \zeta_2}}{N}}
-16\,{\frac {{ \zeta_3}}{N}}
+24\,{\frac {{ \zeta_2}}{{(N+1)}^{2}}}
+8\,{\frac {{ \zeta_3}}{(N+1)}}
-24\,{\frac {{ S_2} ( N+1 ) }{{(N+1)}^{2}}}
\nonumber\\[2ex]&&
-8\,{\frac {{ S_3} ( N+1 ) }{(N+1)}} 
+16\,{\frac {{ S_3} ( N ) }{N}}
-32\,{\frac {{ \beta^{(1)}} ( N+1) }{{N}^{2}}}
+16\,{\frac {{ \beta^{(1)}} ( N+1 ) }{N}}
-80\,{\frac {{ \beta^{(1)}} ( N+2 ) }{{(N+1)}^{2}}}
\nonumber\\[2ex]&&
+16\,{\frac {{ \beta^{(1)}} ( N+2) }{(N+1)}}
-16\,{\frac {{ \beta^{(2)}} ( N+1 ) }{N}}
-8\,{\frac {{ \beta^{(2)}} ( N+2 ) }{(N+1)}}
-96\,{\frac {{ A_3} ( N+1 ) }{(N+1) }}
-192\,{\frac {{ A_3} ( N ) }{N}}
\nonumber\\[2ex]&&
+96\,{\frac {{ \beta} ( N+2) { \zeta_2}}{(N+1)}}
+8\,{\frac {{ A_{18}} ( N+1 ) }{(N+1)}}
-16 \,{\frac {{ A_{18}} ( N ) }{N}}
+ \Bigg( -\,{16 \over N^2}
+\,{8 \over N}\Bigg) { S_2} ( N ) 
+ \Bigg( 16\,{\frac {{ \zeta_2}}{N}}
\nonumber\\[2ex]&&
+192\,{ \frac {{ \beta^{(1)}} ( N+1 ) }{N}}
-16\,{\frac {{ S_2} ( N ) }{N }} \Bigg) { S_1} ( N ) 
+ \Bigg( 96\,{\frac {{ \beta^{(1)}} ( N+2) }{(N+1)}}
+8\,{\frac {{ S_2} ( N+1 ) }{(N+1)}}
\nonumber\\[2ex]&&
-8\,{\frac {{ \zeta_2}}{(N+1)}} \Bigg) { S_1} ( N+1 )  \Bigg] 
\\[2ex] &&
\delta \Delta_{gg}^{(2),C_A}=
 {N_c^2 \over N_c^2-1}\Bigg[ -\,{4 \over N^2}
+\,{101 \over 3 N}
-\,{124 \over 3 (N+1)^2}
-\,{22 \over (N+2)^2}
-\,{101 \over 3 (N+2)}
\nonumber\\[2ex]&&
+\,{8 \over 3 N^3}
-{\frac {56}{3 (N+1)^3}}
-\,{28 \over 3 (N+2)^3}
-16\,{\frac {{ \beta} ( N+1 ) { \zeta_2 }}{N}}
-32\,{\frac {{ S_1} ( N+1 ) { \beta^{(1)}} ( N+2 ) }{(N+1) }}
\nonumber\\[2ex]&&
-8\,{\frac {{ \zeta_3}}{N}}
+16\,{\frac {{ \zeta_3}}{(N+1)}}
-8\,{\frac {{ \zeta_3}}{(N+2)}}
-16\,{\frac {{ S_3} ( N+1 ) }{(N+1)}}
+8\,{\frac {{ S_3 } ( N+2 ) }{(N+2)}}
+8\,{\frac {{ S_3} ( N ) }{N}}
\nonumber\\[2ex]&&
+{16 \over 3}\,{ \frac {{ \beta^{(1)}} ( N+1 ) }{N}}
+\,{\frac {{ 32 ~\beta^{(1)}} ( N+2 ) }{3 (N+1)}}
-4\,{\frac {{ \beta^{(2)}} ( N+1 ) }{N}}
-8\,{ \frac {{ \beta^{(2)}} ( N+2 ) }{(N+1)}}
-4\,{\frac {{ \beta^{(2)}} ( 3+N) }{(N+2)}}
\nonumber\\[2ex]&&
+32\,{\frac {{ A_3} ( N+1 ) }{(N+1)}}
+16\,{\frac {{ A_3} ( N+2 ) }{(N+2)}}
+16\,{\frac {{ A_3} ( N ) }{N}} 
-16\,{\frac {{ S_1} ( N+2 ) { \beta^{(1)}} ( 3+N ) }{(N+2)}}
\nonumber\\[2ex]&&
-16 \,{\frac {{ S_1} ( N ) { \beta^{(1)}} ( N+1 ) }{N}}
-32\,{\frac { { \beta} ( N+2 ) { \zeta_2}}{(N+1)}}
-16\,{\frac {{ \beta} ( 3+N) { \zeta_2}}{(N+2)}}
+{16 \over 3}\,{\frac {{ \beta^{(1)}} ( 3+N ) }{(N+2)}}\Bigg]
\\[2ex]
&&\delta \Delta_{qq,C \overline E}^{(2)}=
\delta \Delta_{qq,D \overline F}^{(2)}=
\delta \Delta_{\overline q\overline q,C \overline E}^{(2)}=
\delta \Delta_{\overline q\overline q,D \overline F}^{(2)}
=\Delta_{qq,C \overline E}^{(2)}
\\[2ex]
&&\delta \Delta_{qq,C \overline F}^{(2)}=
\delta \Delta_{qq,D \overline E}^{(2)}=
\delta \Delta_{\overline q\overline q,C \overline F}^{(2)}=
\delta \Delta_{\overline q\overline q,D \overline E}^{(2)}
=\Delta_{qq,C \overline F}^{(2)}
\\[2ex]&&
\delta \Delta_{gg}^{(2),C_F}=
-\,{55 \over N^2}
+\,{70 \over N}
+\,{72 \over (N+1)^2}
-\,{62 \over (N+1)}
-\,{27 \over (N+2)^2}
-\,{8 \over (N+2)}
\nonumber\\[2ex]&&
+\,{28 \over N^3}
+\,{12 \over (N+1)^3}
-\,{12 \over N^4}
-\,{32 \over (N+1)^4}
+16\,{\frac {{ \beta} ( N+1 ) { \zeta_2}}{N}}
-\,{16 \over (N+2)^4}
\nonumber\\[2ex]&&
+ \Bigg( 16\,\frac{1}{(N+1)^{2}}
+\,{16 \over (N+1)} \Bigg) { S_2} ( N+1 ) 
+ \Bigg( \,{4 \over N^2}
-\,{16 \over N} \Bigg) { S_2} ( N ) 
+ \Bigg( -\,{2 \over N^2}
-\,{8 \over (N+1)^2}
\nonumber\\[2ex]&&
+\,{8 \over N}
-\,{8 \over (N+1)} \Bigg) \ln^2\left({Q^2 \over \mu^2}\right )
-4\,{\frac {{ \zeta_2}}{{N}^{2}}}
+16\, {\frac {{ \zeta_2}}{N}}
+8\,{\frac {{ \zeta_3}}{N}}
-16\,{\frac {{ \zeta_2}}{{(N+1) }^{2}}}
-16\,{\frac {{ \zeta_2}}{(N+1)}}
\nonumber\\[2ex]&&
-16\,{\frac {{ \zeta_3}}{(N+1)}}
+8\,{\frac {{ \zeta_3}}{(N+2)}}
+16\,{\frac {{ S_3} ( N+1 ) }{(N+1)}}
-8\,{\frac {{ S_3} ( N+2 ) }{(N+2)}}
-8\,{\frac {{ S_3} ( N ) }{N}}
-8\,{ \frac {{ \beta^{(1)}} ( N+1 ) }{{N}^{2}}}
\nonumber\\[2ex]&&
+8\,{\frac {{ \beta^{(1)}} ( N+1) }{N}}
-16\,{\frac {{ \beta^{(1)}} ( N+2 ) }{{(N+1)}^{2}}}
+8\,{\frac { { \beta^{(1)}} ( N+2 ) }{(N+1)}}
-16\,{\frac {{ \beta^{(1)}} ( 3+N ) }{ {(N+2)}^{2}}}
+4\,{\frac {{ \beta^{(2)}} ( N+1 ) }{N}}
\nonumber\\[2ex]&&
+8\,{\frac {{ \beta^{(2)}} ( N+2 ) }{(N+1)}}
+4\,{\frac {{ \beta^{(2)}} ( 3+N ) }{(N+2)}}
-32 \,{\frac {{ A_3} ( N+1 ) }{(N+1)}}
-16\,{\frac {{ A_3} ( N+2) }{(N+2)}}
-16\,{\frac {{ A_3} ( N ) }{N}}
\nonumber\\[2ex]&&
+32\,{\frac {{ \beta} ( N+2 ) { \zeta_2}}{(N+1)}}
+16\,{\frac {{ \beta} ( 3+N) { \zeta_2}}{(N+2)}}
+ \Bigg( \,{86 \over N}
+\,{8 \over N^3}
-\,{36 \over N^2}
+16\,{ \frac {{ \beta^{(1)}} ( N+1 ) }{N}} \Bigg) { S_1} ( N ) 
\nonumber\\[2ex]&&
+ \Bigg( 16\,{\frac {{ \beta^{(1)}} ( 3+N ) }{(N+2)}}
+\,{18 \over (N+2)}\Bigg) { S_1} ( N+2 ) 
+ \Bigg( -\,{32 \over (N+1)^2}
-\,{32 \over (N+1)}\Bigg) { S_1^2} ( N+1 ) 
\nonumber\\[2ex]&&
+ \Bigg( -\,{8 \over N^2}+\,{32 \over N} \Bigg) { S_1^2} ( N ) 
+ \Bigg(  \left( \,{32 \over (N+1)}
+\,{32 \over (N+1)^2}\right) { S_1} ( N+1 ) 
+ \left( \,{8 \over N^2}\right.
\nonumber\\[2ex]&& \left.
-\,{32 \over N} \right) { S_1} ( N ) 
-\,{16 \over (N+1)^2}
-\,{16 \over (N+1)^3}
+\,{18 \over N^2}
+\,{52 \over (N+1)}
-\,{4 \over N^3}
-\,{43 \over N}
\nonumber\\[2ex]&&
-\,{9 \over (N+2)} \Bigg) { \ln\left({Q^2 \over \mu^2}\right )}
+ \Bigg( \,{32 \over (N+1)^3}
+\,{32 \over (N+1)^2}
-\,{104 \over (N+1)}
\nonumber\\[2ex]&&
+32\,{\frac {{ \beta^{(1)}} ( N+2 ) }{(N+ 1)}} \Bigg) { S_1} ( N+1 )~. 
\end{eqnarray}
\newpage
{\hspace*{-6mm}\large \bf A.3. Scalar Higgs Coefficient Functions}

\vspace{2mm}\noindent
In this appendix we present the Mellin transforms for the 
coefficient functions
of the hadronic Higgs--production cross section in the heavy--mass limit. 
We used the same notation as in Refs.~\cite{HGS}.

\begin{eqnarray}
\Delta_{ab}=\Delta_{ab}^{(0)}+a_s \Delta_{ab}^{(1)}+a_s^2 \Delta_{ab}^{(2)}
\end{eqnarray}


\newpage
{\hspace*{-6mm}\large \bf A.4 Pseudo-scalar Higgs Coefficient Functions}

\vspace{2mm}\noindent
In this appendix we present the Mellin transforms of the 
coefficient functions
of the hadronic pseudo--scalar Higgs--boson cross section in the 
heavy--mass limit.    We used the same
notation as in  Refs.~\cite{HGS}.

\begin{eqnarray}
\Delta_{ab,A-H}=\Delta_{ab,A-H}^{(0)}+a_s \Delta_{ab,A-H}^{(1)}
+a_s^2 \Delta_{ab,A-H}^{(2)}
\end{eqnarray}

\begin{eqnarray}
&&\Delta_{gg,A-H}^{(0)}=
0
\\[2ex] &&
\Delta_{gg,A-H}^{(1)}=
+8\,{ C_A}
\\[2ex] &&
\Delta_{gq,A-H}^{(1)}=
0
\\[2ex] &&
\Delta_{q \bar q,A-H}^{(1)}=
0
\\[2ex] &&
\Delta_{gg,A-H}^{(2),C_A^2}=
 \Bigg( -{\frac {20}{3}}
-\,{64 \over (N+3)}
+\,{64 \over (N+2)}
-\,{128 \over (N+1)}
-64\,{ S_1} (N-1)  \Bigg) { \ln\left({Q^2 \over \mu^2}\right )}
\nonumber\\[2ex]&&
+{\frac {215}{3}}
-\,{32 \over N^2}
+\,{44 \over 3 N}
-\,{32 \over (N+1)^3}
-\,{440 \over 3 (N+1)^2}
+128\,{\frac {{ S_1} ( N+3 ) }{(N+3)}}
+\,{232 \over 3 (N+1)}
\nonumber\\[2ex]&&
+\,{64 \over (N+2)^2} 
+\,{176 \over 3 (N+3)}
-\,{452 \over 3 (N+2)}
-\,{64 \over (N+3)^2}
+64\, { S_1^2} (N-1) 
\nonumber\\[2ex]&&
+256\,{\frac {{ S_1} ( N+1 ) }{(N+1)}}
+ 128\,{ \zeta_2}
-128\,{\frac {{ S_1} ( N+2 ) }{(N+2)}}
\\[2ex] &&
\Delta_{gg,A-H}^{(2),C_A T_f n_f}=
-{4 \over 3}
+\,{16 \over 3 N}
-\,{32 \over 3 (N+1)^2}
+\,{16 \over 3 (N+1)}
- \,{32 \over 3 (N+2)}
\nonumber\\[2ex]&&
+{\frac {32}{3}}\,{ O_2}\,{ L_t}
-{16 \over 3}\,{ O_2}
-{8 \over 3}\,{ \ln\left({Q^2 \over \mu^2}\right )}
\\[2ex] &&
\Delta_{gg,A-H}^{(2),C_F T_f n_f}=
-50
-\,{16 \over N}
+\,{32 \over (N+1)^3}
+\,{32 \over (N+1)}
-\,{16 \over (N+2)}
-8\,{ \ln\left({Q^2 \over \mu^2}\right )}
\\[2ex] &&
\Delta_{qg,A-H}^{(2),C_F^2}=
-12\,{\frac {{ O_2}}{N}}
-\,{4 \over N}
+\,{16 \over (N+1)^3}
+\,{24 \over (N+1)^2}
+\,{16 \over (N+1)}
+12\,{\frac {{ O_2}}{(N+2)}}
\nonumber\\[2ex]&&
-\,{12 \over (N+2)}
\\[2ex] &&
\Delta_{qg,A-H}^{(2),C_A C_F}=
 \Bigg( \,{32 \over N}
-\,{32 \over (N+1)}
+\,{16 \over (N+2)} \Bigg) { \ln\left({Q^2 \over \mu^2}\right )}
+\,{4 \over (N+2)}
+\,{60 \over N}
\nonumber\\[2ex]&&
-\,{32 \over (N+1)^3}
-\,{80 \over (N+1)^2}
-\,{48 \over (N+1)}
+\,{16 \over (N+2)^2}
-64\,{\frac {{ S_1} ( N ) }{N}}
+64\,{\frac {{ S_1} ( N+1) }{(N+1)}}
\nonumber\\[2ex]&&
-32\,{\frac {{ S_1} ( N+2 ) }{(N+2)}}
\\[2ex] &&
\Delta_{qg,A-H}^{(2),C_F T_f n_f}=
0
\\[2ex] &&
\Delta_{q_1 q_2,A-H}^{(2),C_F^2}=
-\,{32 \over N^2}
+\,{88 \over N}
-\,{32 \over (N+1)^3}
-\,{48 \over (N+1)^2}
-\,{96 \over (N+1)}
+\,{8 \over (N+2)}
\\[2ex] &&
\Delta_{q q,A-H}^{(2),C_A C_F^2}=
\,{32 \over N}
-\,{32 \over (N+1)^3}
-\,{32 \over (N+1)^2}
-\,{32 \over (N+1)}
\\[2ex] &&
\Delta_{q q,A-H}^{(2),C_F^3}=
-\,{64 \over N}
+\,{64 \over (N+1)^3}
+\,{64 \over (N+1)^2}
+\,{64 \over (N+1)}
\\[2ex] &&
\Delta_{q \bar q,A-H}^{(2),C_A C_F^2}=
\,{8 \over 3 N}
+\,{32 \over (N+1)^3}
-\,{80 \over 3 (N+1)^2}
+\,{56 \over 3 (N+2)}
-\,{64 \over 3 (N+3)}
\\[2ex] &&
\Delta_{q \bar q,A-H}^{(2),C_F^3}=
-48\,{\frac {{ O_2}}{N}}
+\,{64 \over N}-\,{32 \over (N+1)^3}
+\,{32 \over (N+1)^2}
+96\,{ \frac {{ O_2}}{(N+1)}}
-\,{160 \over (N+1)}
\nonumber\\[2ex]&&
-48\,{\frac {{ O_2}}{(N+2)}}
+\,{96 \over (N+2)}
\\[2ex] &&
\Delta_{q \bar q,A-H}^{(2),C_F^2 T_f n_f}=
-\,{32 \over 3 N}
+\,{64 \over 3 (N+1)^2}
+\,{32 \over 3 (N+2)}~,
\\[2ex] 
\end{eqnarray}
where
\begin{eqnarray}
O_2=1 \quad \quad L_t=\ln\left({\mu_R^2 \over m_t^2}\right)~. 
\end{eqnarray}

%% file: paper.bbl
\begin{thebibliography}{99} 
%
\bibitem{OCCAM} 
William of Occam, {\sf Quadlibeta}, Book V, (1324). 
%
\bibitem{HMAS} 
R.K. Ellis, I. Hinchliffe, M. Soldate, J.J. van der Bij, Nucl. Phys. 
{\bf B297} (1988) 221; \\
U. Baur, E. Glover, Nucl. Phys. {\bf B339} (1990) 38;\\
D. Graudenz, M. Spira, P. Zerwas, Phys. Rev. Lett. {\bf 70} (1993) 1372;\\
M. Spira, A. Djouadi, D. Graudenz, P. Zerwas, Phys. Lett. {\bf B318} 
(1993) 347;
Nucl. Phys. {\bf B453} (1995) 17.
%
\bibitem{ABDEL}
A. Djouadi, {\tt hep-ph/0503172, hep-ph/0503173.}   
%
\bibitem{DAWS}
S. Dawson, Nucl. Phys. {\bf B359} (1991) 283;
A. Djouadi, M. Spira, P. Zerwas, Phys. Lett. {\bf B264} (1991) 440
%
\bibitem{DY} 
T. Matsuura, S.C. van der Marck, and W.L. van Neerven, Nucl. Phys. {\bf B319} (1989) 570;\\ 
R. Hamberg, W.L. van Neerven, and T. Matsuura, Nucl.~Phys. {\bf B359} 
(1991) 343;\\ 
V.~Ravindran, J.~Smith and W.~L.~van Neerven, Nucl.\ Phys.\ {\bf B682} (2004) 421. 
%
\bibitem{HGS} 
S. Catani, D. de Florian, and M. Grazzini, JHEP (2001) 0105:025;\\ 
R.V. Harlander, W.B. Kilgore, Phys. Rev. {\bf D64} (2001) 013015; Phys. Rev. Lett. {\bf 88} (2002) 201801; 
JHEP (2002) 0210:017;\\ 
C. Anastasiou and K. Melnikov, Nucl. Phys. {\bf B646} (2002) 220;\\ 
V.~Ravindran, J.~Smith, and W.~L.~van Neerven, Nucl.\ Phys.\ {\bf B665} 
(2003) 325; {\bf B704} (2005) 332.\\
J. Smith and W.L. van Neerven, {\tt hep-ph/0501098};\\  
C. Anastasiou, K. Melnikov, and F. Petriello,
{\tt hep-ph/0501130}.
%
\bibitem{JK} 
J. Bl\"umlein and S. Kurth, Phys. Rev. {\bf D60} (1999) 014018. 
%
\bibitem{TEV} 
Tevatron Electroweak Working Group, D0 collab., {\tt 
hep-ex/0404010}.
%
\bibitem{LHC} 
CMS collab., Technical Proposal, report CERN/LHCC/94-38, ATLAS Coll., 
ATLAS Detector and Physics Performance: Technical Design 
Report, Vol.~2, report CERN/LHCC/99-15 (1999). 
%
\bibitem{anNNLO} 
S. Moch, J.A.M. Vermaseren, and A. Vogt, 
Nucl. Phys. {\bf B688} (2004) 101; {\bf B691} (2004) 129.
%
\bibitem{nielsen} 
N. Nielsen, Nova Acta Leopold. {\bf XC} (1909) 121;\\ 
S. K\"olbig, Siam J. Math. Anal. {\bf 17} (1986) 1232;\\ 
L. Lewin, {\sf Dilogarithms and Associated Functions} 
(Macdonald, London, 1958);\\ 
{\sf Polylogarithms and Associated Functions} (North Holland, New York, 1981). 
%
\bibitem{JB3} 
J. Bl\"umlein, Comput. Phys. Commun. {\bf 133} (2000) 76. 
%
\bibitem{JBSM} 
J. Bl\"umlein and S.-O. Moch, {\tt hep-ph/0503188}. 
%
\bibitem{JVE} 
J.A.M. Vermaseren, Int. J. Mod. Phys. {\bf A14} (1999) 2037. 
%
\bibitem{euler} 
L. Euler, Novi Comm. Acad. Sci Petropolitanae {\bf 1} (1775) 140;\\ 
R.L. Graham, D.E. Knuth, and O. Patashnik, {\sf Concrete Mathematics}, 
(Addison-Wesley, Reading/MA, 1994). 
%
\bibitem{GIRG} 
J.M. Borwein and R. Girgensohn, Electron. J. Combinatorics {\bf 3} (1996) 
R23 (Appendix by D.J. Broadhurst). 
%
\bibitem{JB1} J. Bl\"umlein, Comput. Phys. Commun. {\bf 159} (2004) 19. 
%
\bibitem{JB04} 
J. Bl\"umlein,  Nucl. Phys. {\bf B} Proc. Suppl. {\bf 135} 
(2004) 225, {\tt hep-ph/0407044}; DESY 04--064. 
%
\bibitem{JBVR1} 
J. Bl\"umlein and V. Ravindran, Nucl. Phys. {\bf B} Proc. 
Suppl. {\bf 135} (2004) 24, {\tt hep-ph/0407045}. 
%
\bibitem{JBSM1} 
J. Bl\"umlein and S. Moch, DESY 05--008. 
%
\bibitem{JBVR2} 
J. Bl\"umlein and V. Ravindran, in preparation. 
%
\bibitem{JV} 
J. Bl\"umlein and A. Vogt, Phys. Lett. {\bf B370}  (1996) 149; {\bf B 386} (1996) 350;
Acta Phys. Pol. {\bf B 27} (1996) 1309; Phys. Rev. {\bf D 57} (1998) 1; 
{\bf D58} 
(1998) 014020;\\
J. Bl\"umlein, V.~Ravindran, W.L. van Neerven and A. Vogt, 
{\tt hep-ph/9806368};\\ 
R.K. Ellis, Z. Kunszt, and E. Levin, Nucl. Phys. {\bf B420} (1994) 517; 
E~:~{\bf B433} (1995) 498(E). 
%
\bibitem{JBSA} 
S. Alekhin and J. Bl\"umlein, Phys. Lett. {\bf B594} (2004) 299. 
%
\bibitem{mellin} 
H. Mellin, Acta Math. {\bf 25} (1902) 139. 
%
\bibitem{JBHK1} J. Bl\"umlein and H. Kawamura, {\tt hep-ph/0409289} 
and in preparation. 
%
\bibitem{HMAS1} 
M. Kr\"amer, E. Laenen, M. Spira, Nucl. Phys. B511 (1998) 523;\\
K.G. Chetyrkin, B.A. Kniehl, and M. Steinhauser, Phys. Rev. Lett. {\bf 79}
(1997) 353;\\
K.G. Chetyrkin, B.A. Kniehl, M. Steinhauser, and W.A. Bardeen, Nucl. 
Phys. {\bf B535} (1998) 3.
%
\bibitem{NIELSHB} 
N. Nielsen, {\sf Handbuch der Theorie der Gammafunktion}, (Teubner, 
Leipzig, 1906).
%
\bibitem{carlson} 
E. Carlson, Thesis, Univ. Uppsala, 1914, E.C.~Titchmarsh, 
{\sf Theory of Functions}, (Oxford University Press, Oxford, 1939), 
Chap.9.5. 
\end{thebibliography}
